\documentclass[letterpaper]{article} 
\usepackage{aaai2027}  
\usepackage[hyphens]{url}  
\usepackage{graphicx} 
\usepackage{natbib}  
\usepackage{caption} 
\usepackage{algorithm}
\usepackage{algorithmic}

\usepackage{tabularx}
\usepackage{capt-of}

\usepackage{newfloat}
\usepackage{listings}
\DeclareCaptionStyle{ruled}{labelfont=normalfont,labelsep=colon,strut=off} 
\floatstyle{ruled}
\newfloat{listing}{tb}{lst}{}
\floatname{listing}{Listing}

\usepackage{booktabs}

\usepackage{amssymb}
\usepackage{threeparttable}
\usepackage{graphicx}
\usepackage{multirow}
\usepackage{tabularx}
\usepackage{enumitem}
\usepackage{makecell}
\usepackage{caption}
\usepackage{subcaption}
\usepackage{amsmath}

\def\modelname{InitGen}

\newcommand{\squishlist}{
   \begin{list}{$\bullet$}
    { \setlength{\itemsep}{0pt}      \setlength{\parsep}{2pt}
      \setlength{\topsep}{1pt}       \setlength{\partopsep}{0pt}
      \setlength{\leftmargin}{1em} \setlength{\labelwidth}{1em}
      \setlength{\labelsep}{0.5em} } }

\newcommand{\squishend}{
    \end{list}  }

\title{{\modelname}: Candidate Generation for Interaction Initiation in Intelligent Assistants}

\author {
    Ruize Shi\textsuperscript{\rm 1,*},
    Jinhua Chen\textsuperscript{\rm 2},
    Hong Huang\textsuperscript{\rm 1,$\dagger$},
    Ziniu Chen\textsuperscript{\rm 2},
    Ruike Zhang\textsuperscript{\rm 2},
    Jianxun Shi\textsuperscript{\rm 2},\\
    Yitao Chen\textsuperscript{\rm 1},
    Rui Zhang\textsuperscript{\rm 1}
}
\affiliations {
    \textsuperscript{\rm 1}Huazhong University of Science and Technology\\
    \textsuperscript{\rm 2}OPPO AI Center\\
    \{rzshi, honghuang, u202210910\}@hust.edu.cn, \{chenjinhua1, chenziniu1, zhangruike, shijianxun\}@oppo.com, rayteam@yeah.net
}

\begin{document}

\maketitle

\renewcommand{\thefootnote}{*}
\footnotetext[1]{This work was done during the internship at OPPO AI Center.}
\renewcommand{\thefootnote}{$\dagger$}
\footnotetext[2]{Hong Huang is the corresponding author. Ruize Shi, Hong Huang, and Yitao Chen are affiliated with the National Engineering Research Center for Big Data Technology and System, Services Computing Technology and System Lab, Cluster and Grid Computing Lab, School of Computer Science and Technology, Huazhong University of Science and Technology.}
\renewcommand{\thefootnote}{\arabic{footnote}}

\begin{abstract}
Interaction initiation refers to presenting multiple candidate queries when a user opens an intelligent assistant before expressing any intent for the current session. In production, candidate generation incorporates dynamic context and produces all candidates within a strict latency budget. Learning from user feedback is also difficult since the generator usually produces more candidates than are finally displayed. After downstream filtering and ranking, only a subset is exposed to users, so the observed feedback is partial and cannot be reliably assigned to individual queries. We present {\modelname}, a framework for candidate generation in interaction initiation. {\modelname} jointly generates a set of candidate queries and aligns it with user feedback through weighted preference optimization. The sample weights are derived from user activity and downstream ranking scores. The activity weight reduces the dominance of highly active users during training, while the ranking score is used as a practical estimate of the reliability of the observed feedback. {\modelname} also uses a rolling data window to incorporate recent interactions into periodic model updates. In an online A/B test against a strong production baseline, {\modelname} improves the click-through rate from 0.95\% to 1.61\%, corresponding to a relative improvement of 69.1\%, and increases query exposure by 17.9\% under the same traffic allocation. {\modelname} generates the complete candidate set within 180\,ms and has been fully deployed in OPPO's Xiaobu Assistant, which serves over 150 million monthly active users.
\end{abstract}


\section{Introduction}

\begin{figure}[t]
    \centering
    \includegraphics[width=0.99\linewidth]{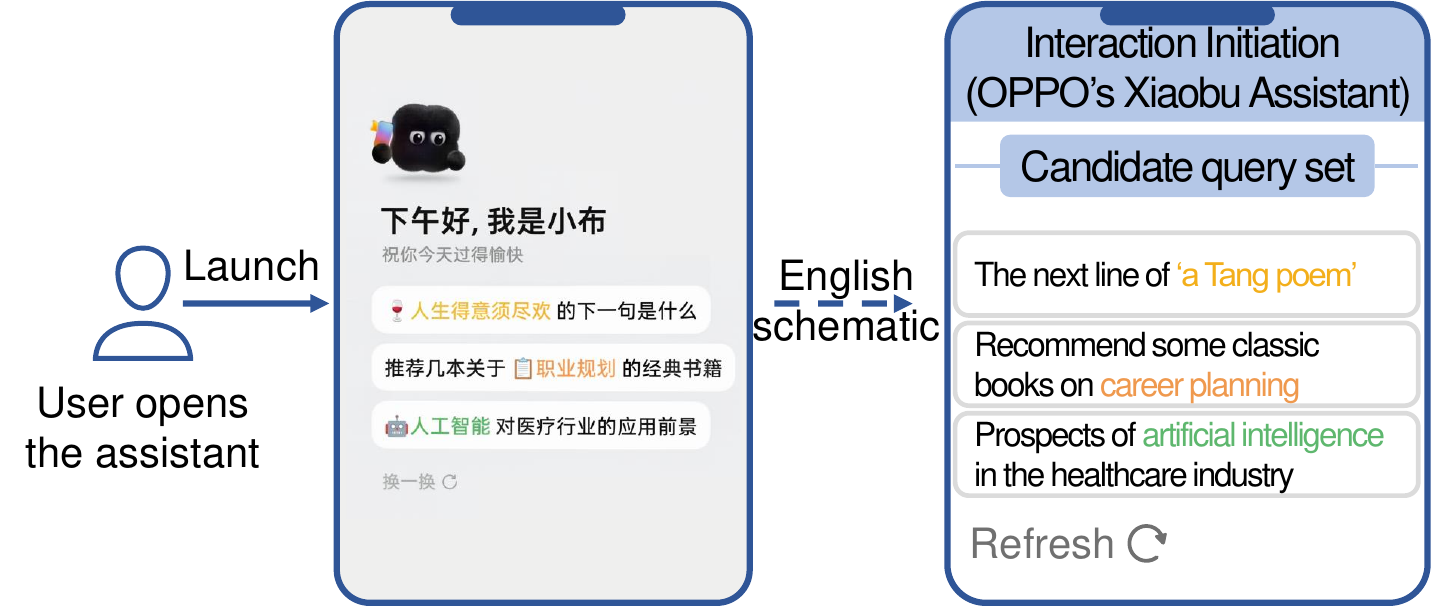}
    \caption{An example interface of interaction initiation in OPPO's Xiaobu Assistant on smartphones. When a user opens the assistant, the interface presents a set of candidate queries to guide subsequent interactions. The user can also refresh the displayed candidates, which triggers the generation of a new candidate set under the same latency requirement.}
    \label{fig:interactioninitiation}
\end{figure}

Intelligent assistants are widely used across mobile devices, smart speakers, and conversational platforms \cite{intelligentassistants1,intelligentassistants2}. When a user opens an assistant before entering a query or expressing an intent for the current session, the interface may proactively present a small set of candidate interactions to guide subsequent engagement, as illustrated in Figure \ref{fig:interactioninitiation}. These candidates are natural language interaction entries and may take the form of queries, instructions, or questions. For simplicity, we refer to each entry as a candidate query and to the process of presenting such candidates at assistant launch as {\em interaction initiation}. Interaction initiation provides users with immediate entry points for subsequent interaction and can influence their early engagement with the assistant.

Despite its practical importance, interaction initiation has only recently begun to receive attention as a distinct generation setting \cite{icebreaker}. Table~\ref{tab:setting_comparison} compares interaction initiation with related problem settings. Many related settings rely on an observed query or dialogue context, select from a predefined candidate space, or receive feedback at the same granularity as their output \cite{rssurvey2,irsurvey1}. In our production setting, interaction initiation generates multiple open-text candidates before the user expresses an intent for the current session, without a predefined candidate pool, and learns from feedback observed only for the subset selected by downstream modules.

\begin{table*}[t]
\centering
    \resizebox{0.99\linewidth}{!}{
    \begin{tabular}{c|cccccc}
        \toprule
        Setting & Query/prefix & Dialogue context & Fixed pool & Open-text generation & Multiple candidates & Matched feedback granularity \\
        \midrule
        Retrieval & $\checkmark$ & $\times$ & $\checkmark$ & $\times$ & $\checkmark$ & $\checkmark$ \\
        Recommendation & $\times$ & $\times$ & $\checkmark$ & $\times$ & $\checkmark$ & $\checkmark$ \\
        Generative query suggestion & $\checkmark$ & $\times$ & $\times$ & $\checkmark$ & $\checkmark$ & $\checkmark$ \\
        Proactive conversational guidance & $\times$ & $\checkmark$ & $\times$ & $\checkmark$ & $\times$ & $\checkmark$ \\
        {\bf Interaction initiation} & $\times$ & $\times$ & $\times$ & $\checkmark$ & $\checkmark$ & $\times$ \\
        \bottomrule
    \end{tabular}
    }
    \caption{Comparison with related problem settings. $\checkmark$ and $\times$ indicate whether a property is typically present. Matched feedback granularity indicates whether observable feedback is available at the same granularity as the model output used for optimization.}
    \label{tab:setting_comparison}
\end{table*}

Under this setting, interaction initiation presents three main challenges. (1) {\em Unknown session intent.} Historical interactions may reveal a user's long-term preferences, but the user has not yet expressed an intent for the current session. This distinguishes interaction initiation from methods driven by an observed query, dialogue context, or inferred session intent \cite{llm4isr,onesug}. (2) {\em Strict online latency.} Users may refresh the displayed candidates at any time, and candidate generation incorporates dynamic context such as location and user profile attributes. In our production setting, the complete candidate set is generated within 180\,ms. Generating queries independently, even in parallel, can also produce repeated or highly similar candidates since the queries are not coordinated during generation. (3) {\em Partial and delayed feedback.} The generator usually produces more candidate queries than are finally displayed. The generated candidates pass through downstream filtering, selection, and ranking, and only a subset is eventually exposed to users. Feedback is therefore observed only after downstream processing and only for the exposed candidates. This makes it difficult to derive reliable labels or preference relations for individual generated queries, unlike settings where feedback is directly associated with individual outputs \cite{rssurvey1}.

These challenges motivate a generative formulation \cite{llmsurvey1} for interaction initiation. At assistant launch, the generator receives user history and context, but no explicit query or fixed candidate pool. Generative models can directly construct natural language candidates that reflect historical preferences while covering different possible intents \cite{agent4rec,grsurvey1,grsurvey2}. Producing the complete candidate set in one generation request also allows the candidates to be coordinated without invoking the model separately for each query. More importantly, the learning objective should match the available feedback. Since user responses are observed only after downstream processing and cannot be reliably decomposed across all generated candidates, optimization is performed over the generated query set rather than over individual queries.

Based on this formulation, we present {\modelname}, a framework for candidate generation deployed in the interaction initiation pipeline of OPPO's Xiaobu Assistant. {\modelname} generates a set of candidate queries jointly in one generation request. The model is initialized through {\em supervised fine tuning} (SFT) \cite{sft} on historical interaction data and then aligned with observed feedback using {\em Kahneman-Tversky Optimization} (KTO) \cite{kto} at the query set level. To account for differences among training samples, {\modelname} derives sample weights from user activity and downstream ranking scores. The activity weight reduces the dominance of highly active users during training, while the ranking score provides a practical estimate of the reliability of the observed feedback. {\modelname} further uses a rolling window of recent interaction data for periodic model updates. These updates are performed offline and do not add latency to online candidate generation. We evaluate {\modelname} through online A/B tests in a large-scale production environment. The complete framework improves both user engagement and the exposure of generated candidates while satisfying the latency requirement of interaction initiation. Our contributions can be summarized as follows:
\squishlist
    \item We study interaction initiation as a candidate generation setting with unknown session intent, strict latency requirements, and feedback observed only after downstream selection and exposure.
    \item We present {\modelname}, a deployed candidate generation component that jointly produces query sets, aligns them using user activity and ranking scores, and periodically incorporates recent data.
    \item We evaluate {\modelname} in OPPO's Xiaobu Assistant against a standard KTO production baseline. It achieves relative improvements of 69.1\% in {\em click-through rate} (CTR) and 17.9\% in exposure, generates each candidate set within 180\,ms, and serves over 150 million monthly active users.
\squishend

\begin{figure*}[t]
    \centering
    \includegraphics[width=0.99\textwidth]{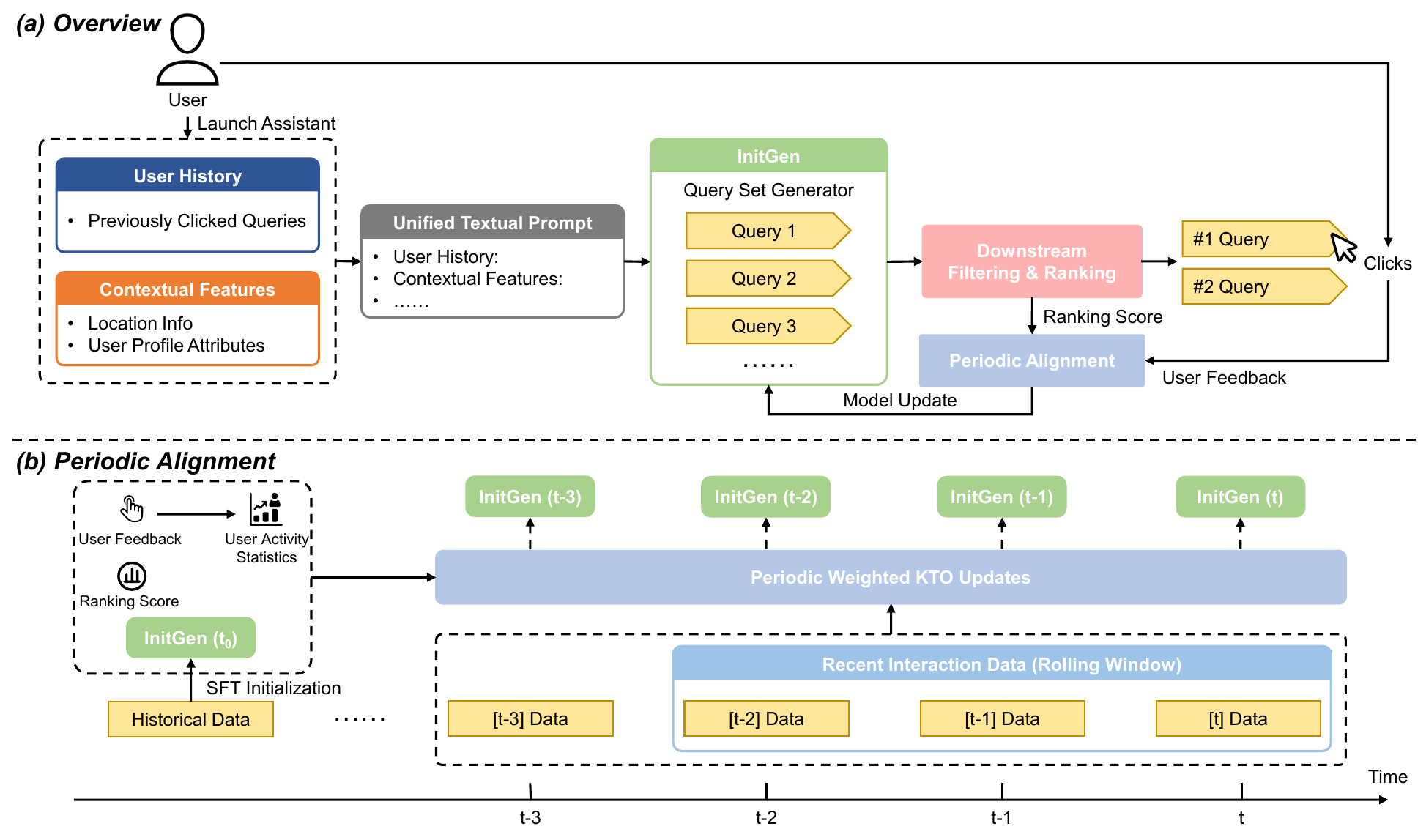}
    \caption{Overview of {\modelname} in the interaction initiation pipeline. (a) Online workflow, where {\modelname} generates a set of candidate queries and interacts with downstream filtering and ranking modules. (b) Offline periodic alignment, where the model is initialized with SFT and periodically updated using weighted KTO and a rolling window of recent interaction data.}
    \label{fig:model}
\end{figure*}

\section{The Proposed Method}

We now describe {\modelname} as illustrated in Figure~\ref{fig:model}. In production, {\modelname} generates candidate queries, while downstream filtering and ranking determine which are displayed. {\modelname} generates a query set in one request under a strict latency budget and is periodically updated with feedback collected after downstream processing and exposure. We first formulate the interaction initiation problem and its practical constraints. We then introduce the joint generation of candidate queries, the weighted alignment strategy, and the rolling update strategy used to incorporate recent interaction data.

\subsection{Problem Formulation}

{\noindent\bf Input.} At the moment of interaction initiation, the generator receives the user's historical interaction data, such as previously clicked queries, together with contextual information, including geographic information and user profile attributes. No current query or query prefix is available, and the user's intent for the current session remains unknown.

{\noindent\bf Output.} The generator produces a set of $K$ natural language candidate queries, denoted by $\mathcal{Q} = \{q_{1}, q_{2}, \ldots, q_{K}\}$. Each candidate can serve as an entry point for subsequent interaction. In practice, $K$ is determined by product requirements and downstream processing capacity and is typically larger than the number of available UI slots. The additional candidates allow downstream filtering and ranking modules to select a subset for display. Unlike recommendation or retrieval systems that select items from a predefined candidate space, interaction initiation generates open-text candidates that collectively cover multiple plausible user intents.

{\noindent\bf Latency constraint.} Candidate generation operates under a strict latency requirement since users may refresh the displayed queries at any time, and contextual information may change between requests. In our production setting, the complete candidate set is generated within 180\,ms. This requirement favors generating all candidates in a single request rather than invoking the model separately for each query. Joint generation also enables coordination among candidates and helps reduce repeated or highly similar outputs.

{\noindent\bf Supervision and feedback.} The available feedback is partial and delayed. The generated candidates pass through downstream filtering and ranking modules, and only a subset is exposed to the user. Consequently, feedback is observed only for exposed candidates after downstream processing. A click on one exposed query does not establish a preference ordering over the complete generated set. Unexposed queries receive no feedback, while exposed but unclicked queries may be affected by display position or competition from other candidates and thus cannot be reliably treated as negatives.

A generation request with no exposed query provides no observable feedback and is excluded from alignment. For requests with at least one exposed query, the observed click outcome is aggregated at the query set level. This produces a set-level alignment signal without assigning potentially unreliable labels or pairwise preferences to individual queries.

{\noindent\bf Problem definition.} Given historical interactions and contextual information, interaction initiation aims to learn a model to generate an effective set of candidate queries $\mathcal{Q}$. The generated set should cover uncertain user intents, meet online latency constraints, and support alignment from partial feedback observed after downstream filtering and ranking.

\subsection{Joint Query Set Generation}\label{sec:generation}

To support interaction initiation under strict latency constraints, we formulate candidate generation at the query set level. Given the input information, the model produces $K$ candidate queries within one generation request rather than generating or scoring each query independently. The candidates are generated together so that the resulting set can cover different possible user intents while limiting redundancy.

{\modelname} uses historical interaction data, such as previously clicked queries, to capture long-term user preferences \cite{srsurvey}. It also incorporates contextual information provided by upstream systems, including geographic information and user profile attributes. These signals are represented in textual form and included in the model input, allowing the generated queries to reflect both user preferences and the current context. As illustrated in Figure~\ref{fig:model}(a), historical interactions and contextual information are combined into a unified textual prompt, from which {\modelname} generates the complete candidate set.

To generate multiple candidates together, {\modelname} represents all $K$ queries as one textual output, with individual queries separated by newline delimiters. Although the output is represented as an ordered sequence for autoregressive generation, it is treated as a candidate set by the downstream pipeline. During generation, each query is conditioned on the shared input and the previously generated content in the same output. This allows the model to coordinate the candidates within one generation process rather than producing them through separate model requests. Such joint generation helps reduce repeated or highly similar queries and supports broader coverage of uncertain user intents.

{\modelname} is initialized from a pretrained {\em large language model} (LLM) and adapted through SFT on historical interaction data. The SFT data use the same output format, with candidate queries separated by newline delimiters. This initialization establishes the joint generation behavior and provides the starting point for subsequent alignment with feedback observed after downstream processing.

\begin{figure}[t]
    \centering
    \includegraphics[width=0.95\linewidth]{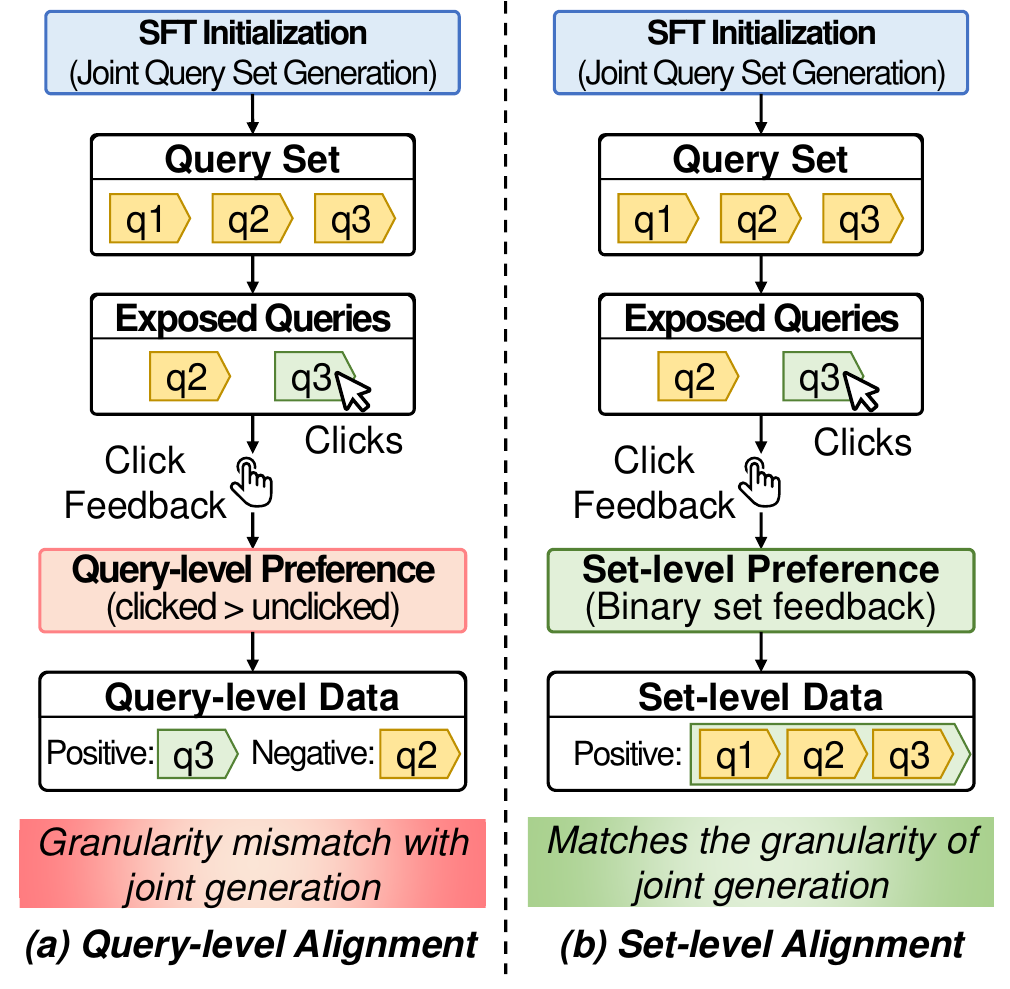}
    \caption{Illustration of alignment granularity. (a) Query level alignment derives labels for individual exposed queries from click feedback, which can introduce unreliable preferences and creates a granularity mismatch with joint query set generation. (b) Query set alignment assigns a binary outcome to the complete generated set without assigning labels to individual queries, matching the granularity of generation.}
    \label{fig:alignment}
\end{figure}

\subsection{Query Set Alignment with Weighted Preference Signals}\label{sec:alignment}

After a query set is generated, the feedback used for alignment is observed only after downstream filtering, ranking, and exposure. Only a subset of the generated queries may be displayed, and unexposed queries receive no user feedback regardless of their quality. The observed outcome is therefore partial and indirect rather than an explicit relevance signal for every generated query.

Figure~\ref{fig:alignment} compares query level and set level alignment under partial feedback. {\modelname} jointly generates all candidate queries as one serialized output, whereas query level alignment derives training signals for individual exposed queries, labeling clicked queries as positive and unclicked queries as negative. Such labels are unreliable since unclicked queries may be affected by display position or competition from other candidates, while unexposed queries receive no feedback. Query level optimization also treats jointly generated candidates as independent decisions, creating a mismatch between the units of generation and optimization.

To avoid this mismatch, {\modelname} treats the complete generated query set as the unit of alignment. Let $\mathcal{E}(\mathcal{Q}) \subseteq \mathcal{Q}$ denote the subset of generated queries exposed to the user. We construct alignment samples only for query sets with $\mathcal{E}(\mathcal{Q}) \neq \emptyset$. The binary outcome of each retained query set is defined as:
\begin{equation}
y(\mathcal{Q}) =
\begin{cases}
1, & \text{if at least one query in $\mathcal{E}(\mathcal{Q})$ is clicked},\\
0, & \text{if no query in $\mathcal{E}(\mathcal{Q})$ is clicked}.
\end{cases}
\end{equation}
A query set with $y(\mathcal{Q})=1$ is considered desirable, while a query set with $y(\mathcal{Q})=0$ is considered undesirable. Query sets with no exposed queries are excluded since they provide no observable user feedback. The label is assigned to the serialized query set as a whole and does not imply that every query in a desirable set is individually positive.

{\modelname} adopts KTO since it learns directly from desirable and undesirable outcomes without requiring explicit preference pairs. Each serialized query set is treated as one completion in the KTO objective. This formulation matches the granularity of joint query set generation and allows the model to learn from binary outcomes when reliable labels for individual queries are unavailable.

Although query set alignment matches the available feedback, individual records carry unequal amounts of evidence. Feedback from highly active users accounts for a disproportionate share of the interaction data \cite{longtailsurvey}. In addition, the confidence associated with an observed outcome varies with the support assigned by the downstream ranking model. {\modelname} accounts for these differences through weights derived from user activity and ranking scores.

First, users are grouped into coarse activity buckets based on their historical interaction frequency. The user activity weight $w_u$ is defined as:
\begin{equation}
w_u =
\sigma\left(\mathrm{CTR}_{avg}-\mathrm{CTR}_{group}\right)+0.5,
\end{equation}
where $\mathrm{CTR}_{avg}$ is the average CTR across all users, $\mathrm{CTR}_{group}$ is the average CTR of the corresponding activity group, and $\sigma(\cdot)$ is the sigmoid function. CTR values in this equation are represented in percentage units. The weight equals one when the group CTR matches the overall CTR. Groups with lower CTR receive weights greater than one, while groups with higher CTR receive weights below one. This reduces the dominance of highly active user groups while keeping the contribution of each record bounded.

Second, {\modelname} uses the downstream ranking score as a practical estimate of the confidence of the observed feedback. The ranking score is not used as a reward or as an objective for increasing ranking scores or exposure. It only determines how strongly an observed outcome contributes to alignment.

The production logs record ranking scores for individual exposed queries. Before alignment, we aggregate records from each generation request into one training sample. For each query set $\mathcal{Q}$ with $\mathcal{E}(\mathcal{Q}) \neq \emptyset$, the ranking signal is the mean score of its exposed queries. This aggregation ensures that each generated query set contributes once to the training objective, regardless of the number of its exposed queries.

Let $s_q$ denote the normalized ranking score of an exposed query $q$. The mean ranking score of a query set is defined as:
\begin{equation}
\bar{s}(\mathcal{Q}) =
\frac{1}{|\mathcal{E}(\mathcal{Q})|}
\sum_{q \in \mathcal{E}(\mathcal{Q})} s_q,
\end{equation}
and the ranking-based weight is then defined as:
\begin{equation}
w_r(\mathcal{Q}) = \bar{s}(\mathcal{Q}) + c,
\end{equation}
where $c$ is determined by the range of the score normalization so that $w_r$ has a range comparable to $w_u$ and is centered at 1. For example, $c$ can be set to 0.5 when ranking scores are normalized to $(0,1)$. A higher mean ranking score indicates stronger downstream support for the exposed queries and therefore assigns greater weight to the observed set outcome, whether desirable or undesirable. Thus, $w_r$ adjusts only the confidence of the alignment signal without changing its label or directly optimizing ranking scores or exposure.

For a training sample associated with query set $\mathcal{Q}$, the final weight is defined as:
\begin{equation}
w(\mathcal{Q}) =
\sqrt{w_u \cdot w_r(\mathcal{Q})}.
\end{equation}
The geometric mean combines the two signals while limiting the influence of either weight on the final objective.

The resulting objective is termed {\em weighted KTO} (WKTO) and is defined as follows:
\begin{equation}
\mathcal{L}_{WKTO}
=
\sum_{\mathcal{Q}\in\mathcal{D}}
w(\mathcal{Q})
\,
\mathcal{L}_{KTO}
\left(\mathcal{Q},y(\mathcal{Q})\right),
\end{equation}
where $\mathcal{D}$ denotes retained query sets with at least one exposed query, and $\mathcal{L}_{KTO}\left(\mathcal{Q},y(\mathcal{Q})\right)$ is the standard KTO loss.

By treating the complete query set as the completion and the unit of preference labeling, {\modelname} maintains consistency between joint generation and alignment. User activity controls the imbalance across user groups, while the mean ranking score controls the strength of the observed evidence. Importantly, ranking scores only scale the alignment signal and do not serve as direct objectives for ranking or exposure.

\subsection{Periodic Alignment with a Rolling Data Window}

Interaction initiation operates in an evolving production environment, where user behavior, available content, and downstream ranking policies may change over time. A model aligned on a fixed collection of historical interactions may therefore become less responsive to recent usage patterns. {\modelname} is periodically updated with newly observed interaction data to adapt to these changes.

One option is to retain all previously collected preference data during successive updates. However, older feedback may reflect outdated user interests or earlier pipeline versions. Repeated optimization over an expanding data collection may also increase the influence of frequently observed patterns and reduce the contribution of recent interactions. More broadly, prior work has shown that repeated training on model influenced data can amplify existing distributional biases and degrade generation quality \cite{modelscollapse}. Although our setting differs from recursive training on generated data, this observation motivates controlling the age and composition of the feedback used for periodic alignment.

As illustrated in Figure~\ref{fig:model}(b), {\modelname} is first initialized through SFT on historical interaction data. At each subsequent update, WKTO is applied to the current model using interaction records from a rolling window of recent data. Records outside the window are excluded from the current update. The window moves forward as new interaction data become available, allowing each update to reflect recent user behavior and the current downstream environment.

The rolling window serves two practical purposes. First, it allows the model to incorporate recent interaction patterns without repeating SFT or retraining on the complete historical dataset. Second, it limits the influence of stale feedback and prevents the update data from growing indefinitely. The SFT initialized model provides the starting point for periodic alignment, while each update adapts the current model using a bounded collection of recent preference signals.

All alignment updates are performed offline. The updated model is deployed only after training and validation, so periodic alignment does not add latency to online candidate generation. The deployed model continues to generate the complete candidate set within the same latency budget.

Overall, periodic alignment with a rolling data window provides a practical way to incorporate recent interaction data while keeping the update process compatible with production serving requirements.

\section{Experiments}

\subsection{Experimental Setup}

We evaluate {\modelname} in the production interaction initiation pipeline of OPPO's Xiaobu Assistant on smartphones, which serves over 150 million monthly active users. When a user opens the assistant, the candidate generation component produces a set of queries based on user history and contextual information within a strict latency budget. Details of data privacy protection are provided in the Ethical Statement.

{\noindent\bf Evaluation setting.}
We use online A/B testing as the primary evaluation. Interaction initiation occurs before the user expresses an intent for the current session, and click feedback is available only for candidates displayed after downstream filtering and ranking. These characteristics make reliable offline relevance labels and preference comparisons difficult to construct. Online experiments directly measure the effectiveness of generated candidates in the production environment.

{\noindent\bf Metrics.}
The primary metric is CTR, defined as the number of user clicks divided by the number of exposed candidate queries. CTR measures how often the exposed candidates provide effective entry points for subsequent interaction. We also report query exposure, defined as the number of generated candidate queries displayed after downstream filtering and ranking. Under the same traffic allocation, query exposure reflects how frequently generated candidates are accepted by the existing downstream pipeline.

{\noindent\bf Baseline.} To satisfy the strict latency requirements of online interaction initiation, all compared models use Qwen2.5-1.5B \cite{qwen2.5} as the backbone. We compare {\modelname} with a strong baseline trained using SFT followed by standard KTO. Both models perform alignment at the query set level and use the same downstream filtering and ranking pipeline.

{\noindent\bf Training and inference details.} The learning rates are $2\times10^{-5}$ for SFT and $1\times10^{-6}$ for KTO and WKTO. The SFT model serves as the reference model during alignment, and the KL regularization parameter $\beta$ is 0.1. We set the temperature to 0.7, top-$p$ to 0.8, and top-$k$ to 20. Each request jointly generates five candidate queries, with at most three displayed after downstream filtering and ranking. For rolling updates, the model is updated daily using newly collected production interactions. Further practical lessons from training and deployment are provided in Appendix A.1.

\begin{table}[t]
\centering
    \resizebox{0.99\linewidth}{!}{
    \begin{tabular}{l|ccc}
        \toprule
        Method & Exposure & Clicks & CTR \\
        \midrule
        Baseline (SFT + standard KTO) & 3,810,832 & 36,263 & 0.95\% \\
        Ours ({\modelname}) & {\bf 4,492,970} & {\bf 72,328} & {\bf 1.61\%} \\
        \midrule
        $\Delta$ (Relative lift) & +17.9\%   & +99.5\% & +69.1\% \\
        \bottomrule
    \end{tabular}
    }
    \caption{Online study under identical traffic allocation.}
    \label{tab:onlinestudy}
\end{table}

\begin{figure}[t]
    \centering
    \includegraphics[width=0.9\linewidth]{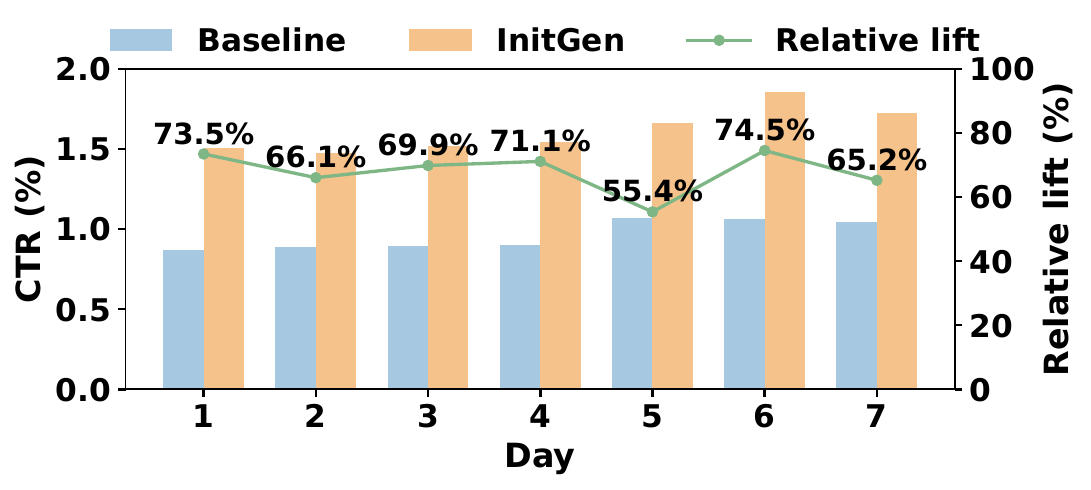}
    \caption{Daily CTR and relative lift of {\modelname} against the production baseline over one week.}
    \label{fig:oneweekctr}
\end{figure}

{\noindent\bf A/B testing protocol.}
Online experiments are conducted through controlled traffic allocation in the production system. For each comparison, users are randomly assigned to control and treatment groups with identical traffic allocation. The downstream filtering and ranking pipeline and the user interface remain unchanged between the two groups.

The complete {\modelname} framework and the production baseline are evaluated within the same experimental window. Due to production deployment and user experience constraints, the component variants are evaluated through separate pairwise A/B tests during staged deployment, with each new variant compared with its immediate predecessor. Each experiment runs for one week or longer, excluding holidays and periods of abnormal traffic.

\subsection{Online Study}

We first evaluate the online performance of {\modelname} against the production baseline. Both are evaluated in the same window with identical traffic allocation and an unchanged downstream pipeline. Table~\ref{tab:onlinestudy} presents the aggregated results.

{\modelname} increases query exposure from 3.81 million to 4.49 million, corresponding to a relative improvement of 17.9\%. With the downstream pipeline unchanged, this increase shows that the query sets generated by {\modelname} provide more candidates that satisfy the existing downstream criteria and reach the user interface.

We can also see that CTR increases from 0.95\% to 1.61\%, yielding a relative lift of 69.1\%. The larger CTR gain indicates that the overall benefit comes primarily from stronger engagement with the candidates that are eventually displayed, rather than simply from increasing the number of exposures. Together, the improvements in downstream exposure and user response result in a 99.5\% increase in total clicks.

Figure~\ref{fig:oneweekctr} shows daily CTR over the experiment. {\modelname} outperforms the production baseline on every day. The CTR varies across the experimental period, but the improvement remains positive under different daily traffic conditions. In particular, {\modelname} retains a clear gain on days when the baseline CTR is relatively high, showing that the improvement is not limited to periods in which the baseline performs poorly.

Overall, these results show that {\modelname} is superior to the production baseline. More generated candidates are exposed, and exposed candidates are more likely to initiate interaction.

\begin{table}[t]
\centering
\resizebox{0.99\linewidth}{!}{
\begin{tabular}{c|cccc|c}
\toprule
\multicolumn{5}{c|}{Model} & \multirow{2}{*}{$\Delta$CTR} \\
\cmidrule{1-5}
ID & Context & $w_u$ & $w_r$ & Rolling Update &  \\
\midrule
1 & $\times$ & $\times$ & $\times$ & $\times$ & -- \\
2 & $\checkmark$ & $\times$ & $\times$ & $\times$ & {\bf+11.6\%} \\
3 & $\checkmark$ & $\checkmark$ & $\times$ & $\times$ & {\bf+5.7\%} \\
4 & $\checkmark$ & $\checkmark$ & $\checkmark$ & $\times$ & {\bf+19.6\%} \\
{\modelname} & $\checkmark$ & $\checkmark$ & $\checkmark$ & $\checkmark$ & {\bf+20.1\%} \\
\bottomrule
\end{tabular}
}
\caption{Ablation study of {\modelname}. ID~1 corresponds to the production baseline. Context includes geographic information and user profile attributes. Each subsequent variant is compared with its immediate predecessor in a separate pairwise A/B test, and $\Delta$CTR reports the relative CTR lift for that comparison.}
\label{tab:ablationstudy}
\end{table}

\subsection{Ablation Study}\label{sec:ablationstudy}

We conduct a staged online ablation study of the main components in {\modelname}. Starting from the production baseline based on SFT and standard KTO, each variant adds one component to the immediately preceding model. Due to production deployment and user experience constraints, the variants are evaluated through separate pairwise A/B tests in successive experimental windows. Table~\ref{tab:ablationstudy} reports the relative CTR lift within each pairwise comparison. 

Adding geographic information and user profile attributes available at assistant launch improves CTR by 11.6\%. The improvement shows that these signals help narrow the range of plausible interactions when no explicit intent is available for the current session. They complement historical interactions by providing information about the user's current environment and general interests.

Using the user activity weight $w_u$ further improves CTR by 5.7\%. Interaction logs are unevenly distributed across users, and highly active users contribute a large proportion of the observed feedback. Weighting samples according to user activity reduces this imbalance during preference alignment. The positive result supports accounting for differences in user activity when learning from production feedback.

Adding the ranking score weight $w_r$ yields an additional CTR improvement of 19.6\%. The ranking score is used to control the strength of each observed preference signal according to its estimated confidence. The result indicates that treating all click outcomes equally leaves useful information unused, especially when feedback is observed only after downstream processing. Incorporating $w_r$ allows the alignment process to place greater emphasis on outcomes that receive stronger support from the existing downstream model.

\begin{figure}[t]
    \centering
    \includegraphics[width=0.6\linewidth]{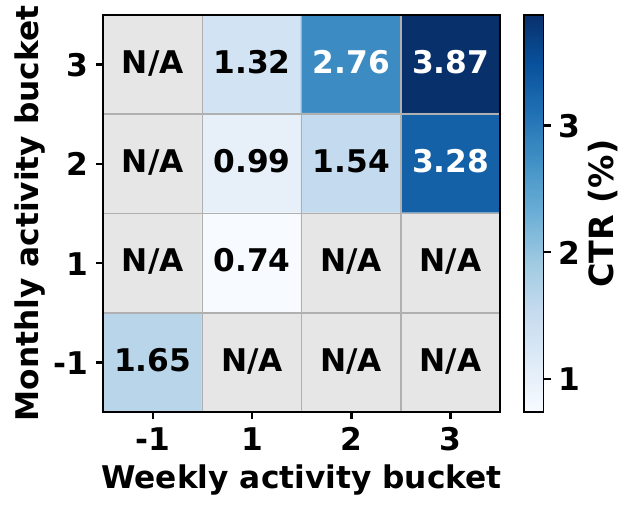}
    \caption{CTR across user activity buckets formed from weekly and monthly activity signals. N/A denotes bucket combinations that do not occur in the production data.}
    \label{fig:userheatmap}
\end{figure}

\begin{table}[t]
\centering
\resizebox{0.99\linewidth}{!}{
\begin{tabular}{c|c|ccc c}
\toprule
User Cohort & Model & Exposure & Clicks & CTR & $\Delta$CTR \\
\midrule
\multirow{2}{*}{High-CTR users}
& ID 2 & 481,005 & 9,692 & 2.01\% & -- \\
& ID 3 & 483,080 & 9,830 & 2.03\% & +1.0\% \\
\midrule
\multirow{2}{*}{Low-CTR users}
& ID 2 & 1,645,727 & 12,876 & 0.78\% & -- \\
& ID 3 & 1,645,089 & 13,906 & 0.85\% & +9.0\% \\
\midrule
\multirow{2}{*}{All users}
& ID 2 & 2,126,732 & 22,568 & 1.06\% & -- \\
& ID 3 & 2,128,169 & 23,736 & 1.12\% & +5.7\% \\
\bottomrule
\end{tabular}
}
\caption{User activity weight $w_u$ study of {\modelname}. ID~2 uses contextual inputs, while ID~3 additionally applies $w_u$. Users are divided into high-CTR and low-CTR cohorts according to whether their historical CTR is above or below the overall average. $\Delta$CTR reports the relative lift of ID~3 over ID~2.}
\label{tab:userweight}
\end{table}

The final variant enables daily rolling updates and improves CTR by a further 20.1\%. This result shows the value of incorporating recently collected interaction data into model alignment. User behavior and the downstream environment evolve over time, and updating the model with a recent data window allows the generator to adapt to these changes without repeating the complete supervised training process.

Overall, the staged comparisons show complementary gains from contextual inputs, weighted preference signals, and rolling updates. Context improves the information available at generation time, the two weighting factors refine how partial feedback contributes to alignment, and rolling updates incorporate recent interaction patterns. We further provide a case study in Appendix A.2 to illustrate how geographic information and user profile attributes affect {\modelname}.

\begin{figure}[t]
    \centering
    \includegraphics[width=0.9\linewidth]{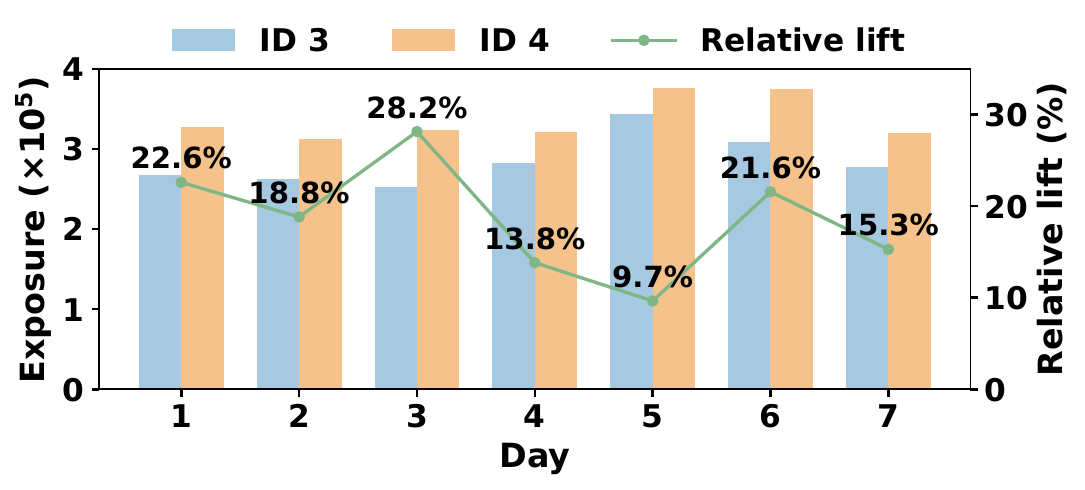}
    \caption{Daily exposure gain of ID~4 (with ranking score weight $w_r$) over one week under the same traffic.}
    \label{fig:scorestudy}
\end{figure}

\subsection{Weighting Study}

We examine the two weighting factors used in WKTO. The ablation study shows that adding the user activity weight $w_u$ improves CTR by 5.7\%, while the ranking score weight $w_r$ brings a further improvement of 19.6\%. We next analyze their effects across user groups and on downstream exposure.

We first study the user activity weight $w_u$. Figure~\ref{fig:userheatmap} shows CTR across groups defined by weekly and monthly activity. Across regular activity buckets, CTR generally increases with user activity. Highly active users thus contribute more records and stronger click signals to the training data, which can cause their feedback to have a disproportionate influence during alignment. The activity weight adjusts the contribution of different user groups to reduce this imbalance.

Table~\ref{tab:userweight} compares ID~2 and ID~3 across user cohorts divided according to whether their historical CTR is above or below the overall average. After introducing $w_u$, the improvement is concentrated among users with lower historical response rates. Their CTR increases from 0.78\% to 0.85\%, corresponding to a relative gain of 9.0\%, while the high CTR cohort remains stable at around 2.0\%. This asymmetric effect raises the overall CTR from 1.06\% to 1.12\% and is consistent with the design of $w_u$, which increases the contribution of user groups with weaker click signals and reduces the dominance of users with stronger historical engagement.

We next study the ranking score weight $w_r$, which adjusts the contribution of each preference record according to the estimated confidence. As shown in Figure~\ref{fig:scorestudy}, ID~4 achieves higher query exposure than ID~3 on every day of the one week experiment, with relative improvements ranging from 9.7\% to 28.2\%. Together with the 19.6\% CTR improvement reported in the ablation study, the results show that $w_r$ improves the quality of preference alignment by placing greater emphasis on more reliable feedback. The resulting query sets are more likely to pass downstream filtering and ranking and receive stronger user response after exposure.

\subsection{Efficiency Study}

We evaluate the serving efficiency of {\modelname} under production traffic. In OPPO's Xiaobu Assistant, the service must return the complete query set within 180\,ms. {\modelname} runs using vLLM~\cite{vllm} on a cluster of 20 NVIDIA A100 GPUs with 80\,GB of memory each.

\begin{figure}[t]
    \centering
    \begin{subfigure}[t]{\linewidth}
        \centering
        \includegraphics[width=0.95\linewidth]{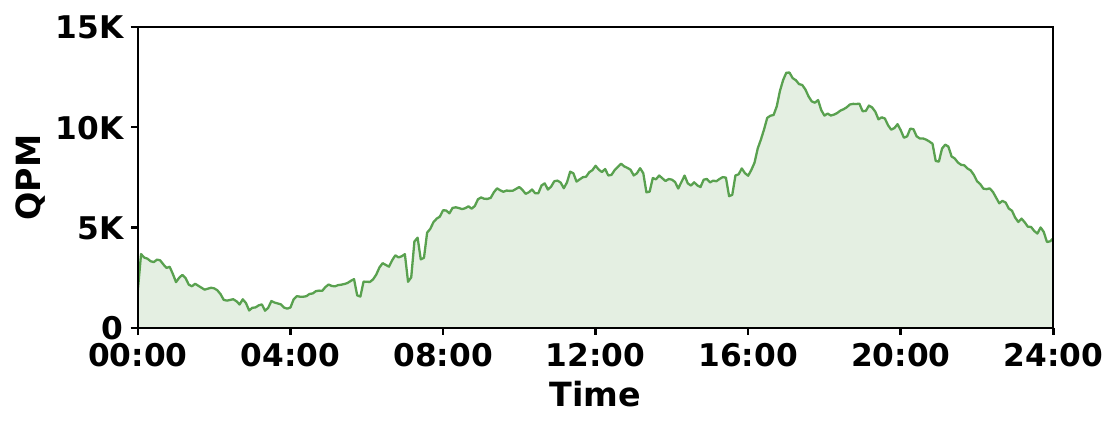}
        \caption{QPM over one day.}
    \end{subfigure}
    
    \begin{subfigure}[t]{\linewidth}
        \centering
        \includegraphics[width=0.95\linewidth]{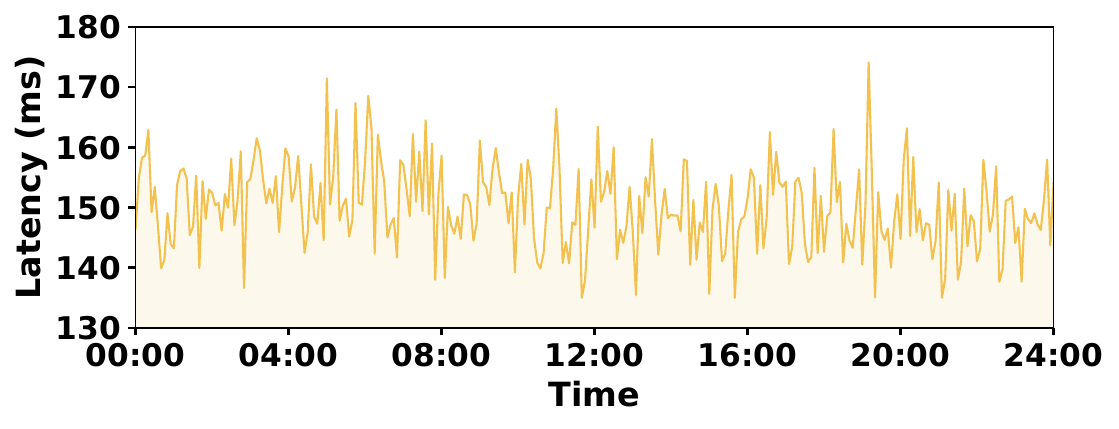}
        \caption{Candidate generation latency over one day.}
    \end{subfigure}

    \caption{Efficiency study of {\modelname}. QPM is the number of query-set generation requests processed per minute. The service handles peak traffic above 12K QPM while keeping candidate-set generation latency below the 180\,ms budget.}
    \label{fig:efficiency}
\end{figure}

Figure~\ref{fig:efficiency} reports QPM and candidate generation latency over one day. The workload varies substantially with user activity, increasing from around 1K QPM during the early morning to more than 12K QPM near the daily peak. Despite this variation, the generation latency remains concentrated around 150\,ms and stays below the 180\,ms budget. In particular, the peak in QPM is not accompanied by a sustained increase in latency, showing that the deployed serving configuration can accommodate daily traffic variation while satisfying the latency requirement. These results demonstrate that joint query set generation is practical under the current production traffic and hardware configuration.

\section{Related Work}

Interaction initiation is related to proactive interaction, recommendation systems, and preference alignment for LLMs. It differs from most of these settings in the information available at generation time, the form of the output, and the feedback observed after deployment.

One relevant line of research examines proactive interaction in search and intelligent systems, where candidate queries or actions are presented before or during an explicit user request \cite{proactive1,proactive2}. Representative settings include query suggestion \cite{querysuggestion,mei2008query,gqs}, proactive guidance \cite{gsftcrl}, and recommendation without an explicit query \cite{drn,ngcf}. Query suggestion typically uses a query or prefix, proactive guidance relies on dialogue context or session signals, and recommendation without an explicit query selects from predefined items or actions \cite{rssurvey4}. Concurrent work considers a similar problem by using a learned model to estimate user interaction and construct preference pairs \cite{icebreaker}. In contrast, \modelname{} learns directly from observed interaction outcomes, without constructing preference pairs, and generates the complete candidate set within a 180 ms production latency budget.

Our work is also related to recommendation systems \cite{rssurvey2,rssurvey3}, including recent generative approaches using LLMs \cite{gpt2,gpt3}. Classical recommendation methods, such as collaborative filtering, typically use users' historical interactions to infer their preferences and rank items from a predefined catalog \cite{matrixrs,ncf}. More recent methods use generation at different stages to produce textual representations, explanations, or intermediate queries for candidate retrieval and item matching \cite{llmrec,dealrec,allmrec}. End-to-end generative recommendation has also emerged \cite{onerec,ega,grsurvey3}. Interaction initiation differs in its output space and feedback process. It generates a small set of natural language queries before the session intent is known, without a predefined item catalog or query pool. After downstream filtering and ranking, only a subset is exposed, so feedback cannot be reliably assigned to every generated query. {\modelname} therefore jointly generates candidates and aligns query sets using aggregated feedback.

From an optimization perspective, {\modelname} is related to LLM alignment. Early alignment methods commonly use reinforcement learning from human feedback \cite{ppo,cobra}, which trains a reward model from preference annotations and then optimizes LLMs through reinforcement learning \cite{rlsurvey1,rlsurvey2}. Recent methods simplify preference alignment through direct objectives such as DPO \cite{dpo}, SimPO \cite{simpo}, and ORPO \cite{orpo}. GRPO \cite{grpo} instead estimates a reinforcement learning baseline from group scores. KTO learns from independently labeled desirable and undesirable outputs without explicit preference pairs \cite{kto}, which suits interaction initiation since click outcomes do not provide reliable pairwise preferences among queries. However, standard KTO does not account for differences among samples, motivating our WKTO design.

\section{Conclusion}
We study interaction initiation as a candidate generation setting for intelligent assistants, where the session intent is unknown and feedback is observed only after downstream processing. We present {\modelname}, a framework that jointly generates a query set in one generation request, aligns the complete set with aggregated click outcomes through WKTO, and incorporates recent interaction data through periodic updates with a rolling data window. In a production A/B test on OPPO's Xiaobu Assistant, {\modelname} improves CTR by 69.1\% from 0.95\% to 1.61\%, and query exposure by 17.9\%. It generates each candidate set within 180\,ms and is deployed in an assistant serving over 150 million monthly active users. Future work could explore offline metrics for generated query quality and their correlation with user engagement.

\section*{Ethical Statement}
All data used for training and evaluation were anonymized before use and contained no direct personal identifiers. The geographic signal used by {\modelname} is limited to coarse city-level information rather than precise location coordinates. User profile attributes consist of high-level interest descriptions generated from previously clicked queries by an existing upstream component and are refreshed periodically. Users can disable personalization through the product settings. Data preparation, model training, evaluation, and online experimentation were conducted entirely on internal company servers. No raw user data were transferred to external services or disclosed outside the company, and only aggregate experimental results are reported.

\section*{Acknowledgments}
This work was supported by the Huazhong University of Science \& Technology--OPPO Research Center for Intelligent System.

\bibliography{aaai2027}

\newpage
\appendix

\setcounter{table}{0}
\setcounter{figure}{0}
\renewcommand{\thetable}{A\arabic{table}}
\renewcommand{\thefigure}{A\arabic{figure}}

\begin{table*}[htb]
\centering
\begin{tabular}{p{0.18\linewidth}|p{0.78\linewidth}}
\toprule
Setting & Content \\
\midrule
\textbf{Input (summary)} &
\textbf{User profile:}
I like recording and organizing information and frequently use note-taking features; I am interested in photography and image processing, especially AI-based photo enhancement; I enjoy writing, particularly travel-related articles. \newline
\textbf{Historical interactions:}
\begin{itemize}[noitemsep,topsep=0pt,leftmargin=*]
\item Will there be strong winds in Shenzhen in the coming days
\item What places are worth visiting in Dongguan
\item Deep thinking: the long-term psychological impact of dreaming about being trapped in a forest
\item Deep thinking: psychological reasons for dreaming about being in a future society
\item How is the weather in Dongguan recently
\item What impact did Zhang Qian's journey to the Western Regions have on later generations
\item What are some famous coral reef regions in the world
\item Weather forecast for Shenzhen over the next three days
\item What is the market outlook for traditional handicrafts
\item How to take impressive photos using everyday objects
\item Try the photo motion blur repair feature
\item Where are good places to shoot night views in Shenzhen
\end{itemize}
\textbf{Current location:} Wuhan, Hubei Province. \\
\midrule
\textbf{No context} &
\begin{itemize}[noitemsep,topsep=0pt,leftmargin=*]
\item Deep thinking: how the popularity of short videos influences teenagers' reading habits
\item What intangible cultural heritage exhibitions are available in Shenzhen recently
\item Where are good places to watch the sunset in Shenzhen on weekends
\item Deep thinking: can dream interpretation predict personal health
\item What are some suitable short-distance weekend trip destinations around Dongguan
\end{itemize} \\
\midrule
\textbf{+ Location} &
\begin{itemize}[noitemsep,topsep=0pt,leftmargin=*]
\item Recommended weekend trip destinations around Wuhan
\item What is the recent air quality index in Wuhan
\item Deep thinking: what do frequent dreams indicate
\item Will it rain in Wuhan tomorrow
\item How to understand the impact of the Silk Road on the modern economy
\end{itemize} \\
\midrule
\textbf{+ Location \& profile} &
\begin{itemize}[noitemsep,topsep=0pt,leftmargin=*]
\item Deep thinking: how short-video platforms influence literary creation
\item Which artistic streets in Wuhan are suitable for photography
\item The significance of coral reef protection for marine ecosystems
\item How AI photo editing improves depth in landscape photos
\item Writing tips for travel-related articles
\end{itemize} \\
\bottomrule
\end{tabular}
\caption{Case study of contextual signals in {\modelname}. All textual content is translated from Chinese into English for presentation.}
\label{tab:casestudy}
\end{table*}

\section{Appendix}

\subsection{A.1 Training and Deployment Lessons}

{\noindent\bf Backbone selection.} The strict latency budget makes backbone selection particularly important. We select Qwen2.5-1.5B since it provides a practical balance between generation quality and serving efficiency. We also evaluate Qwen3-1.7B \cite{qwen3} with thinking mode disabled, but it still frequently exceeds the 180\,ms latency budget under our serving configuration. This result suggests that a newer or moderately larger backbone does not necessarily provide a better production trade-off under strict latency constraints. We additionally observe only limited changes in generation behavior when varying temperature, top-$p$, and top-$k$ within reasonable ranges. This limited sensitivity may be related to the capacity of the compact backbone. We therefore retain the default Qwen2.5 decoding configuration rather than tuning these parameters for individual variants.

{\noindent\bf Joint generation.} We also consider generating candidate queries through independent parallel requests. Since each request is conditioned on the same user information without observing the other generated candidates, we frequently observe duplicate or highly similar queries during system development. After downstream filtering removes redundant or unsuitable candidates, the remaining set may be insufficient to fill the available display slots. Maintaining enough distinct candidates would require generating more queries or issuing additional requests, which repeats prompt processing and increases both serving and downstream processing costs. We therefore generate all candidate queries as one serialized output, allowing the candidates to be coordinated within a single generation request.

{\noindent\bf Alignment granularity and stability.} In our experiments, the Qwen2.5-1.5B backbone is sensitive to the learning rate used for preference alignment. Larger learning rates are associated with unstable generation, motivating the substantially lower learning rate used for KTO and WKTO than for SFT, as reported in the Experimental Setup. We also test DPO with query-level preference pairs on a model initialized through SFT. The resulting models sometimes produce semantically incoherent queries with unintended mixing of Chinese and English, or excessively long outputs that continue until the maximum generation length is reached. This instability may result from the mismatch between the units of supervised initialization and preference alignment, potentially together with the limited model capacity. These observations motivate maintaining the query set as the common output unit for both supervised initialization and subsequent alignment in our implementation.

{\noindent\bf Balance of alignment outcomes.} KTO learns from independently labeled desirable and undesirable outputs and does not require equal numbers of the two outcomes \cite{kto}. However, when training directly on the naturally imbalanced feedback in our setting, we observe generation degeneration similar to that described above, including semantically incoherent queries with mixed Chinese and English and excessively long outputs. Clicks are relatively sparse in interaction initiation, so undesirable query sets substantially outnumber desirable ones. We therefore construct the alignment data with a 1:1 ratio by retaining all available desirable query sets and downsampling undesirable ones accordingly. Consequently, the number of desirable query sets determines the effective amount of data used for each alignment update. Although this behavior may depend on the backbone and data distribution, balanced sampling produces more stable generation in our implementation.

\subsection{A.2 Case Study}
\label{app:casestudy}

We show a case to illustrate how contextual signals affect the query set generated for interaction initiation. We focus on contextual inputs since their influence can be directly observed in individual outputs, while weighted alignment and rolling updates mainly affect the overall generation distribution and are evaluated through the aggregate online studies.

As shown in Table~\ref{tab:casestudy}, without contextual information, the queries mainly follow patterns in the user's historical interactions. Several candidates remain associated with Shenzhen and Dongguan, although the user's current location is Wuhan. The set also includes broader topics related to psychology, culture, and travel. Although these outputs are consistent with the interaction history, some remain less suitable for the current situation since they do not distinguish persistent interests from information tied to previous locations.

Adding the current location shifts the geographic focus toward Wuhan and produces candidates related to local travel, weather, and air quality. At the same time, the model retains broader topics supported by the interaction history, including dreams and historical events. The location signal therefore updates the situational part of the query set without replacing the user's other interests. This allows the candidates to reflect the current environment while covering several possible directions for subsequent interaction.

Further adding profile attributes introduces queries related to photography, AI photo editing, literary creation, and travel writing. Some candidates also combine the current location with longer-term interests, such as photography locations in Wuhan. The example shows that location improves relevance to the current environment, while profile attributes guide longer-term personalization. Together, these signals help {\modelname} balance contextual relevance, persistent user interests, and diversity across the generated query set.

\end{document}